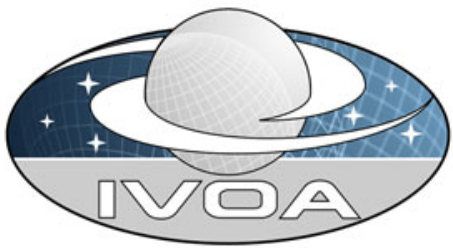

# StandardsRegExt: a VOResource Schema Extension for Describing IVOA Standards
# Version 1.0

## IVOA Recommendation
## 8 May 2012

**Authors:**
   Paul Harrison, Editor
   Douglas Burke
   Ray Plante
   Guy Rixon
   Dave Morris
   and the IVOA Registry Working Group.

## Abstract

This document describes an XML encoding standard for metadata about IVOA standards themselves, referred to as StandardsRegExt. It is intended to allow for the discovery of a standard via an IVOA identifier that refers to the standard. It also allows one to define concepts that are defined by the standard which can themselves be referred to via an IVOA identifier (augmented with a URL fragment identifier). Finally, it can also provide a machine interpretable description of a standard service interface. We describe the general model for the schema and explain its intended use by interoperable registries for discovering resources.

## Status of this document

This document has been produced by the IVOA Registry Working Group.

It has been reviewed by IVOA Members and other interested parties, and has been endorsed by the IVOA Executive Committee as an IVOA Recommendation as 28 April 2012. It is a stable document and may be used as reference material or cited as a normative reference from another document. IVOA's role in making the Recommendation is to draw attention to the specification and to promote its widespread deployment. This enhances the functionality and interoperability inside the Astronomical Community.

Pre-1.0 versions of this document were known as VOStandard.

A list of current IVOA Recommendations and other technical documents can be found at http://www.ivoa.net/Documents/.

## Acknowledgements

This document has been developed with support from the National Science Foundation's Information Technology Research Program under Cooperative Agreement AST0122449 with The Johns Hopkins University, from the UK Particle Physics and Astronomy Research Council (PPARC), and from the European Commission's Seventh Framework Program.

This document contains text lifted verbatim, with small changes, and with substantial changes from the previously published VODataService specification [VDS] (e.g. sections 2.0). This has been done without specific attribution as a means for providing consistency across similar documents. We acknowledge the authors of that document for this text.

## Conformance-related definitions

The words "MUST", "SHALL", "SHOULD", "MAY", "RECOMMENDED", and "OPTIONAL" (in upper or lower case) used in this document are to be interpreted as described in IETF standard, RFC 2119 [RFC 2119].

The **Virtual Observatory (VO)** is a general term for a collection of federated resources that can be used to conduct astronomical research, education, and outreach. The International Virtual Observatory Alliance (IVOA) is a global collaboration of separately funded projects to develop standards and infrastructure that enable VO applications.

XML document **validation** is a software process that checks that an XML document is not only well-formed XML but also conforms to the syntax rules defined by the applicable schema. Typically, when the schema is defined by one or more XML Schema [Schema] documents (see next section), validation refers to checking for conformance to the syntax described in those Schema documents. This document describes additional syntax constraints that cannot be enforced solely by the rules of XML Schema; thus, in this document, use of the term validation includes the extra checks that goes beyond common Schema-aware parsers which ensure conformance with this document.

## Syntax Notation Using XML Schema

The eXtensible Markup Language, or XML, is document syntax for marking textual information with named tags and is defined by the World Wide Web Consortium (W3C) Recommendation, XML 1.0 [XML]. The set of XML tag names and the syntax rules for their use is referred to as the document schema. One way to formally define a schema for XML documents is using the W3C standard known as XML Schema [Schema].

This document defines the StandardsRegExt schema using XML Schema. The full Schema document is listed in Appendix A. Parts of the schema appear within the main sections of this document; however, documentation nodes have been left out for the sake of brevity.

Reference to specific elements and types defined in the VOResource schema include the namespaces prefix, `vr`, as in `vr:Resource` (a type defined in the VOResource schema). Reference to specific elements and types defined in the StandardsRegExt schema include the namespaces prefix, `vstd`, as in `vstd:ServiceStandard` (a type defined in the StandardsRegExt schema). Use of the `vstd` prefix in compliant instance documents is strongly recommended, particularly in the applications that involve IVOA Registries (see [RI], section 3.1.2). Elsewhere, the use is not required.

# Contents

# 1. Introduction

An important goal of the IVOA is to publish standards for services which can interoperate to create a Virtual Observatory (VO). Central to the coordination of these services is the concept of a registry ([RI]) where resources can be described and thus discovered by users and applications in the VO. The standard Resource Metadata for the Virtual Observatory [Hanisch et al. 2004] (hereafter referred to as **RM**) defines metadata terms for services and other discoverable resources. A specific XML encoding of these resources is described in the IVOA standard VOResource: an XML Encoding Schema for Resource Metadata [Plante et al. 2006] (hereafter referred to as **VOResource**). In this schema, support for a standard service protocol is described as a service's *capability*; the associated metadata is contained within the service resource description's `<capability>` element. The specific standard protocol supported is uniquely identified via an attribute of the `<capability>` element called `standardID` whose value is a URI. The VOResource standard does not place a formal validation requirement on the `standardID` other than it be a legal URI; however, it was intended that IVOA-endorsed standards would be represented via an IVOA identifier. As per the IVOA Identifier standard [ID], an IVOA identifier must be registered as a resource in an IVOA-compliant registry.

This document defines a VOResource extension schema called **StandardsRegExt** which allows one to describe a standard and register it with an IVOA registry. By doing so, a unique IVOA identifier becomes "attached" to the standard which can be referred to in other resource descriptions, namely for services that support the standard. Not only does this aid in the unambiguous discovery of compliant service instances but also in validating their descriptions and support for the standard. Another benefit of associating an IVOA identifier with a standard is that it allows registry users who discover services that conform to a particular standard to also discover the document that describes that standard.

StandardsRegExt has two other purposes. First, it allows a service protocol description to communicate specifics about the standard input parameters and output formats specified by the standard. Such a machine-readable description of the interface can assist intelligent portals and applications to build GUI interfaces to standard services and manage workflows built around them. Second, it allows for the definition of small controlled sets of standardized names (referred to as *keys* in this document) which might be used to identify, for example, specific features of a standard protocol (such as supported data transport protocols). By virtue of being defined within the context of a VOResource description, one can refer to the key using a globally unique URI by adding the key name as a URI fragment [URI] onto the IVOA identifier associated with the descriptions.

It is envisaged that StandardsRegExt instances that describe standards endorsed or otherwise in development by the IVOA will be published in the IVOA Registry of Registries [RofR] using the authority identifier [ID], `ivoa.net`. However, other standards, be they ad hoc or endorsed by another body, may be published in any compliant registry.

## 1.1. The Role in the IVOA Architecture

The IVOA Architecture [Arch] provides a high-level view of how IVOA standards work together to connect users and applications with providers of data and services, as depicted in the diagram in Fig. 1.

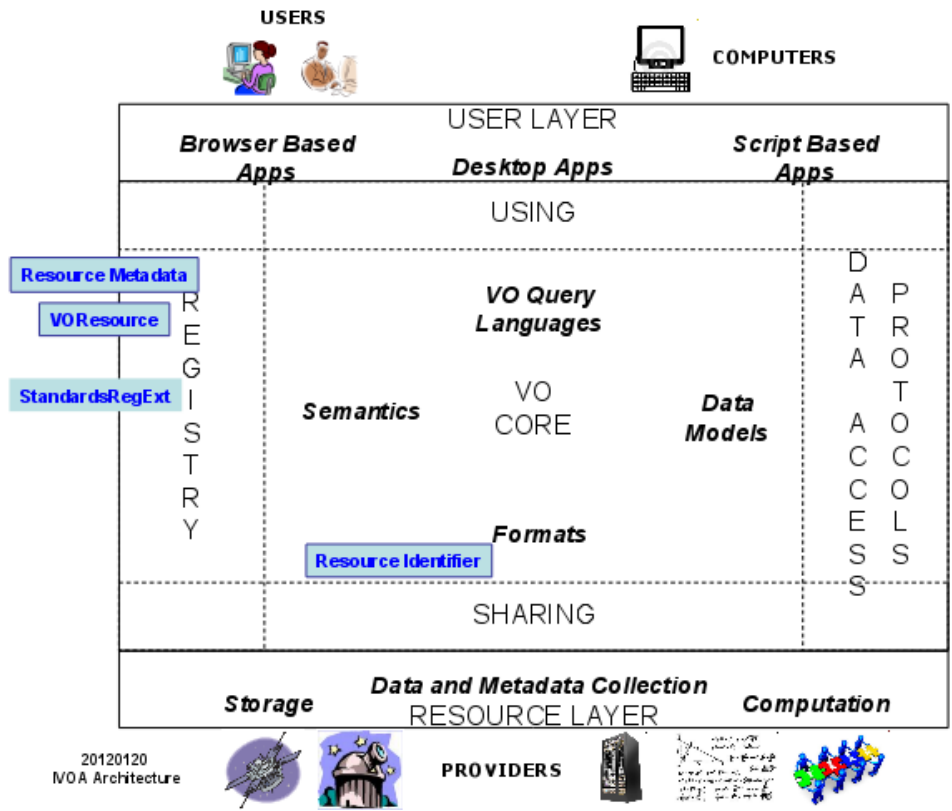

**Figure 1. StandardsRegExt in the IVOA Architecture.** The Registry enables applications in the User Layer to discover archives and services in the Resource Layer. The registry metadata model standards (in blue text and boxes) give structure to the information that enables that discovery. StandardsRegExt defines metadata for describing standards themselves (like those that define the Data Access Protocols).

In this architecture, users can leverage a variety of tools (from the User Layer) to discover archives and services of interest (represented in the Resource Layer); registries provide the means for this discovery. A registry is a repository of descriptions of resources that can be searched based on the metadata in those descriptions. In general, a resource can be more than just archives, data, or services; an IVOA standard, as represented by an IVOA document, can also be a resource. The Resource Metadata standard [RM] defines the core concepts used in the resource descriptions, and the VOResource standard [VOR] defines the XML format. StandardsRegExt is an extension of the VOResource standard that defines extra metadata for describing a standard.

## 2. The StandardsRegExt Data Model

The StandardsRegExt extension in general enables the description of three types of resources:

- a generic standard (specified by an external document)
- a standard specifically defining a service protocol
- a set of related, standardized names called *keys*.

Here's an example of defining a controlled list of computer languages that might be referred to in other descriptions of applications.

**An example of defining a list of programming languages**

```
<ri:Resource xsi:type="vstd:StandardKeyEnumeration" created="2001-12-31T12:00:00"
              updated="2001-12-31T12:00:00" status="active">
    <title>application languages</title>
    <identifier>ivo://ivoa.net/std/application/languages</identifier>
    <curation>
        <publisher>IVOA</publisher>
        <creator>
            <name>IVOA</name>
            <logo>http://www.ivoa.net/icons/ivoa_logo_small.jpg</logo>
        </creator>
        <date role="representative">2006-07-17</date>
        <version>1.0</version>
        <contact>
            <name>IVOA Grid and Web Services WG</name>
            <email>grid@ivoa.net</email>
        </contact>
    </curation>
    <content>
        <subject>IVOA Standard: registry</subject>
        <description>
            This resource defines keys for commonly used computer languages.
        </description>
        <referenceURL>http://www.ivoa.net/twiki/bin/view/IVOA/IvoaResReg</referenceURL>
    </content>
    <key>
        <name>C</name>
        <description>The C programming language</description>
    </key>
    <key>
        <name>CPP</name>
        <description>The C++ programming language</description>
    </key>
    <key>
        <name>CSharp</name>
        <description>The C# programming language</description>
    </key>
    <key>
        <name>FORTRAN</name>
        <description>The FORTRAN programming language</description>
    </key>
    <key>
        <name>Java</name>
        <description>The Java programming language</description>
    </key>
    <key>
        <name>Perl</name>
        <description>The Perl programming language</description>
    </key>
    <key>
        <name>Python</name>
        <description>The Python programming language</description>
    </key>
</ri:Resource>
```

This description defines the meaning behind the following URI, namely the Python language:

```
ivo://ivoa.net/std/application/languages#Python
```

An application can thus refer to, for example, its support for the Python language via this URI. Should other languages become prevalent, the resource description could be updated to add the new names, or a new resource description could be created (with a new IVOA identifier).

## 2.1. The Schema Namespace and Location

The namespace associated with StandardsRegExt extensions is
"http://www.ivoa.net/xml/StandardsRegExt/v1.0". Just like the namespace URI for the VOResource schema, the StandardsRegExt namespace URI can be interpreted as a URL. Resolving it will return the XML Schema document (given in Appendix A) that defines the StandardsRegExt schema.

Authors of VOResource instance documents may choose to provide a location for the VOResource XML Schema document and its extensions using the `xsi:schemaLocation` attribute. While the choice of the location value is the choice of the author, this specification recommends using the StandardsRegExt namespace URI as its location URL (as illustrated in the example above), as in,

```
xsi:schemaLocation="http://www.ivoa.net/xml/StandardsRegExt/v1.0
                    http://www.ivoa.net/xml/StandardsRegExt/v1.0"
```

> **Note:**
> The IVOA Registry Interface standard [RI] actually *requires* that the VOResource records it shares with other registries provide location URLs via `xsi:schemaLocation` for the VOResource schema and all legal extension schemas that are used in the records. This rule would apply to the StandardsRegExt schema.

The prefix, `vstd`, is used by convention as the prefix defined for the StandardsRegExt schema; however, instance documents may use any prefix of the author's choosing. In applications where common use of prefixes is recommended (such as with the Registry Interface specification [RI]), use of the `vstd` prefix is recommended. Note also that in this document, the `vr` prefix is used to label, as shorthand, a type or element name that is defined in the VOResource schema, as in `vr:Resource`.

As recommend by the VOResource standard [VR], the StandardsRegExt schema sets `elementFormDefault="unqualified"`. This means that it is not necessary to qualify element names defined in this schema with a namespace prefix (as there are no global elements defined). The only place it is usually needed is as a qualifier to a StandardsRegExt type name given as the value of an `xsi:type` attribute.

## 2.2. Summary of Metadata Concepts

The StandardsRegExt extension defines three new types of resources. Two are specifically for independently documented standards:

`vstd:Standard`
: This resource describes a general standard (e.g. data model, schema, protocol, etc.). The most important piece of metadata associated with this resource is the `<referenceURL>` (from the core VOResource schema) which should point to the human-readable specification document that defines the standard. It also allows one to provide the recommended version of the standard to use.

`vstd:ServiceStandard`
: This resource type, which extends from `vstd:Standard`, is specifically for describing a standard service protocol (e.g. Simple Cone Search). It differs from `vstd:Standard` in that it also allows one to describe specific constraints on the service interface via its `<interface>` element.

`vstd:StandardKeyEnumeration`
: This resource type allows for the description of a related set of controlled names (referred to as *keys*) and their meanings. While keys can be defined as part of a `vstd:Standard` or `vstd:ServiceStandard` resource, the `vstd:StandardKeyEnumeration` allows a set of key definitions to stand as a resource on its own, regardless of whether it is part of a documented standard or not.

> **Note:**
>> As mentioned above, this standard allows controlled names to be defined either as part of a record of any of the above three types. When such names are being defined as part of an IVOA standard, it is recommended that the `vstd:Standard` or `vstd:ServiceStandard` record corresponding to the IVOA standard document be used to define the names. The `vstd:StandardKeyEnumeration` might be convenient for definitions that are experimental or otherwise are for non-IVOA-based applications.

## 2.3. Defining Enumerations of Identifiers

A common practice when defining metadata is to restrict certain string values to a controlled set of defined names, each with a well-defined meaning. With XML Schema, the controlled set can be enforced by a validating parser (using the `xsd:enumeration` construct [Schema]). One disadvantage of locking in the vocabulary in an XML Schema document is that in order to grow the list of allowed names, a revision of the XML Schema document is required, which can be a disruptive change. To avoid this, it is the practice within VOResource and its extensions to avoid "hard-coded" enumerations in the XML Schema document for sets of defined values that will likely change over time.

The StandardsRegExt schema provides an alternative to XML Schema-based definitions of controlled names. Instead, a controlled list of names, called *standard keys*, can be defined as part of any of the three StandardsRegExt resource types. Updating a resource description is much less disruptive than a Schema document, and as a resource is available via an IVOA-compliant registry, it is still possible for a (non-Schema-based) application to validate the use of the vocabulary.

The StandardsRegExt specification also defines a mapping from a key name to a URI. This allows these keys--and their underlying meaning--to be referenced in a globally unique way in a variety of contexts, not restricted to XML.

A key is defined using the `vstd:StandardKey` type which consists simply of a name and a description. The key is mapped to a URI by attaching the name as the "fragment"--i.e., appending after a pound sign, `#`--to the IVOA identifier for the resource description that defines the key:

  *ivoa-identifier*`#`*key-name*

where *ivoa-identifier* is the resource's IVOA identifier and *key-name* is the name of a key defined in that resource. Consistent with the URI standard [URI], the *key-name* must not contain a pound (#) sign.

For example, we consider a resource description with an IVOA identifier given by `<identifier>` `ivo://ivoa.net/std/QueryProtocol </identifier>` that defines a a key named `case-insensitive`; the URI identifying this key would be:

  `ivo://ivoa.net/std/QueryProtocol#case-insensitive`

This form of defining multiple keys, each with its own mapping to a URI, all in one resource has several advantages:

- The enumerations are naturally grouped under a single registry resource, and so can be retrieved with one registry query and need no further metadata to assert the association.
- The "Dublin core" metadata that is associated with a resource need only be entered once for the whole enumeration, rather than for each member of the enumeration - this saves both curation effort and space in the registry.
- If it is necessary to expand the list of controlled names (or shrink it), it is simple and fairly undisruptive to update the VOResource record.

> **Note:**
> When these enumerations are presented to a user in a GUI it is expected that only the "fragment" part that distinguishes the various members of the enumeration will be used as a choice value, as the full IVO ID is not usually particularly "user-friendly".

Some applications may wish to publish additional metadata associated with a key definition through further extension of VOResource metadata. This can be be done by deriving a new key metadatum type derived by extension from the `vstd:StandardKey`.

# 3. The StandardsRegExt Metadata

## 3.1. Resource Type Extensions

This specification defined three new resource types. As is spelled out in the VOResource specification, a resource description indicates that it refers to one of these types of resources by setting the `xsi:type` attribute to the namespace-qualified type name. Doing so implies that the semantic meaning of that type applies to the resource.

### 3.1.1. Standard

The `vstd:Standard` resource type describes any general standard specification. This typically refers to an IVOA standard but is not limited to such. Generally, the `vstd:Standard` type is intended for standards *other than* standard protocols (which should use the `vstd:ServiceStandard` type instead). It extends the generic `vr:Resource` type as follows.

---

**vstd:Standard Type Schema Definition**

```
<xs:complexType name="Standard" >
  <xs:complexContent >
    <xs:extension base="vr:Resource" >
      <xs:sequence >
        <xs:element name="endorsedVersion" type="vstd:EndorsedVersion"
                    maxOccurs="unbounded" />
        <xs:element name="schema" type="vstd:Schema" minOccurs="0"
                    maxOccurs="unbounded" >
        <xs:element name="deprecated" type="xs:token" minOccurs="0" />
        <xs:element name="key" type="vstd:StandardKey" minOccurs="0"
                    maxOccurs="unbounded" />
      </xs:sequence>
    </xs:extension>
  </xs:complexContent>
</xs:complexType>
```

---

As one of the purposes of this resource type is to enable users to discover the documentation that defines the standard that the resource describes, the `<referenceURL>` should point either to the standard's specification document or to summary information about the standard that can lead one to the specification document.

The `vstd:Standard` resource type adds two metadata terms to the core set:

| vstd:Standard Extension Metadata Elements | |
|---|---|
| **Element** | **Definition** |
| endorsedVersion | *Value type:* a string with optional attributes |
| | *Semantic Meaning:* the version of the standard that is recommended for use. |
| | *Occurrences:* required; multiple occurrences allowed. |
| | *Comments:* More than one version can be listed, indicating that any of these versions are recommended for use. |
| schema | *Value type:* composite vstd:Schema |
| | *Semantic Meaning:* a description and pointer to a schema document defined by this standard. |
| | *Occurrences:* optional; multiple occurrences allowed. |
| | *Comments:* This is most typically an XML Schema, but it need not be strictly. |
| deprecated | *Value type:* string: `xs:token` |
| | *Semantic Meaning:* when present, this element indicates that all versions of the standard are considered deprecated by the publisher. The value is a human-readable explanation for the designation. |
| | *Occurrences:* optional |
| | *Comments:* The explanation should indicate if another standard should be preferred. |
| key | *Value type:* composite `vstd:StandardKey` |
| | *Semantic Meaning:* a defined key associated with this standard. |
| | *Occurrences:* optional; multiple occurrences allowed. |

The child `<key>` elements define terms with special meaning to the standard; see section 3.2.

The purpose of the required `<endorsedVersion>` element is to point potential users of the standard to the version that is most preferred by the standard's publisher. If multiple versions are relevant or in use, multiple elements may be given; in this case, the `use` attribute can further help steer the users to the preferred version.

**vstd:EndorsedVersion Type Schema Definition**

```
<xs:complexType name="EndorsedVersion" >
  <xs:simpleContent >
    <xs:extension base="xs:string" >
      <xs:attribute name="status" default="n/a" >
        <xs:simpleType >
          <xs:restriction base="xs:string" >
            <xs:enumeration value="rec" />
            <xs:enumeration value="pr" />
            <xs:enumeration value="wd" />
            <xs:enumeration value="iwd" />
            <xs:enumeration value="note" />
            <xs:enumeration value="n/a" />
          </xs:restriction>
        </xs:simpleType>
      </xs:attribute>
      <xs:attribute name="use" >
        <xs:simpleType >
          <xs:restriction base="xs:string" >
            <xs:enumeration value="preferred" />
            <xs:enumeration value="deprecated" />
          </xs:restriction>
        </xs:simpleType>
      </xs:attribute>
    </xs:extension>
  </xs:simpleContent>
</xs:complexType>
```

| vstd:EndorsedVersion Attributes | |
|---|---|
| **Attribute** | **Definition** |
| status | *Value type:* string with controlled vocabulary |
| | *Semantic Meaning:* the IVOA status level of this version of the standard. |
| | *Occurrences:* optional |
| | *Allowed Values:* `rec` an IVOA Recommendation |
| | `pr` an IVOA Proposed Recommendation |
| | `wd` an IVOA Working Draft |
| | `iwd` an internal Working Draft of an IVOA Working Group |
| | `note` a published IVOA Note |
| | `n/a` not an IVOA standard or protostandard at this time. |
| | *Default Value:* `n/a` |
| | *Comments:* For values of **rec**, **pr**, **wd**, and **note**, the record's **\<referenceURL\>** element should point to the official specification document in the IVOA Document Repository; if the document does not appear there, these values should not be used. |
| use | *Value type:* string with controlled vocabulary |
| | *Semantic Meaning:* A designation of preference for the version compared to other versions in use. |
| | *Occurrences:* optional |
| | *Allowed Values:* `preferred` the most preferred version. |
| | `deprecated` a version whose use is now discouraged because a newer version is preferred. |

When all versions of the standard are considered deprecated by the resource publisher, the `<deprecated>` child element should appear. The explanation given as a value should mention the standard that the current standard is deprecated by when relevant.

> **Note:**
> An example where the `<deprecated>` element might be used in the VO is in the case of the SkyNode standard [SkyNode]. As of this writing, there are many instances of SkyNode services available in the VO, and where they are used, version 1.01 is endorsed; however, the IVOA has deprecated this standard in favor of the Table Access Protocol [TAP]. Thus, a `vstd:ServiceStandards` record for SkyNode should include a `<deprecated>` element whose content refers viewers to the TAP standard.

The `<schema>` element allows one to list the locations of any schemas defined by the standard thereby making them discoverable as well (just as the specification document is discoverable via the `<referenceURL>` element). It also can provide pointers to example uses of the schemas. Typically (especially for IVOA standards), the schema is an XML schema, and the location points to an XML Schema [Schema] document; however, this is not required. Other schema types and definition formats are allowed.

| vstd:Schema Type Schema Definition |
|---|

```
<xs:complexType name="Schema">
    <xs:sequence>
        <xs:element name="location" type="xs:anyURI"
                    minOccurs="1" maxOccurs="1"/>
        <xs:element name="description" type="xs:token"
                    minOccurs="0" maxOccurs="1" />
        <xs:element name="example" type="xs:anyURI"
                    minOccurs="0" maxOccurs="unbounded" />
    </xs:sequence>
    <xs:attribute name="namespace" type="xs:token" use="required"/>
</xs:complexType>
```

As multiple schemas can be listed in the resource description, the `namespace` attribute provides an identifying label for each `<schema>` element:

| vstd:Schema Attributes | |
|---|---|
| **Attribute** | **Definition** |
| namespace | *Value type:*    string: `xs:token` |
| | *Semantic Meaning:* an identifier for the schema being described. Each instance of this attribute must be unique within the resourse description. |
| | *Occurrences:*    required |
| | *Comments:*    For XML schemas, this should be the schema's namespace URI. Otherwise, it is a unique label to distinguish it from other schemas described in the same resource description. |

The main component of the `<schema>` content is the URL pointing to the schema's definition document, but it can also provide additional information useful for display:

| vstd:Schema Metadata Elements | |
|---|---|
| **Element** | **Definition** |
| location | *Value type:*    URL: `xs:anyURI` |
| | *Semantic Meaning:* A URL pointing to a document that formally defines the schema. |
| | *Occurrences:*    required |
| | *Comments:*    The document should be in a machine-parsable format when applicable. For example, when refering to an XML schema, the document should be an XML Schema or similar document that can be used to validate an instance document. |
| description | *Value type:*    URL: `xs:token` |
| | *Semantic Meaning:* A human-readable description of what the schema defines or is used for. |
| | *Occurrences:*    optional |
| | *Comments:*    A brief description--e.g. one statement--is recommended for display purposes. |
| example | *Value type:*    URL: `xs:anyURI` |
| | *Semantic Meaning:* A URL pointing to a sample document that illustrates the use of the schema. |
| | *Occurrences:*    optional; multiple occurances are allowed. |
| | *Comments:*    When applicable (e.g. XML), the document should be in the format defined by the schema document. |

**An example of a Standard resource that summarizes this specification**

```xml
<?xml version="1.0" encoding="UTF-8"?>
<ri:Resource xsi:type="vstd:Standard" status="active"
             created="2012-02-17T11:15:00" updated="2012-02-17T11:15:00"
             xmlns:ri="http://www.ivoa.net/xml/RegistryInterface/v1.0"
             xmlns:vstd="http://www.ivoa.net/xml/StandardsRegExt/v1.0"
             xmlns:xsi="http://www.w3.org/2001/XMLSchema-instance">
    <title> StandardsRegExt: a VOResource Schema Extension for Describing IVOA Standards </title>
    <shortName> StandardsRegExt </shortName>
    <identifier>   ivo://ivoa.net/std/StandardsRegExt   </identifier>
    <curation>
      <publisher ivo-id="ivo://ivoa.net/IVOA">
        International Virtual Observatory Alliance
      </publisher>
      <creator>
        <name>  IVOA Registry Working Group   </name>
        <logo>http://www.ivoa.net/icons/ivoa_logo_small.jpg</logo>
      </creator>
      <date role="creation"> 2010-05-19 </date>
      <date role="updated">  2012-02-17 </date>
      <version> 1.0 </version>
      <contact>
        <name>   Resource Registry Working Group   </name>
        <email>  registry@ivoa.net      </email>
      </contact>
    </curation>
    <content>
      <subject>    software standard   </subject>
      <subject>   virtual observatory  </subject>
      <description>
        This document describes an XML encoding standard for metadata about
        IVOA standards themselves, referred to as StandardsRegExt.  It is
        intended to allow for the discovery of a standard via an IVOA
        identifier that refers to the standard.  It also allows one to define
        concepts that are defined by the standard which can themselves be
        referred to via an IVOA identifier (augmented with a URL fragment
        identifier).  Finally, it can also provide a machine interpretable
        description of a standard service interface.  We describe the general
        model for the schema and explain its intended use by interoperable
        registries for discovering resources.
      </description>
      <referenceURL>
         http://www.ivoa.net/Documents/StandardsRegExt/
      </referenceURL>
      <type>   Other   </type>
      <contentLevel>   Research   </contentLevel>
    </content>
    <endorsedVersion status="pr"> 1.0 </endorsedVersion>
    <schema namespace="http://www.ivoa.net/xml/StandardsRegExt/v1.0">
      <location>http://www.ivoa.net/xml/StandardsRegExt/v1.0</location>
      <description>
        the VOResource extension XML Schema for registering standards
      </description>
      <example>http://rofr.ivoa.net/examples/StandardsRegExt.xml</example>
      <example>http://rofr.ivoa.net/examples/SIA.xml</example>
      <example>http://rofr.ivoa.net/examples/VOSpace.xml</example>
    </schema>
</ri:Resource>
```

### 3.1.2. ServiceStandard

The `vstd:ServiceStandard` resource type extends `vstd:Standard` to describe more specifically a standard protocol. It adds an `<interface>` element to allow the interface defined by the standard to be described in a machine-readable way. Its type is defined to be `vr:Interface`, which is defined in the VOResource schema [VR].

```
vstd:ServiceStandard Type Schema Definition

<xs:complexType name="ServiceStandard" >
  <xs:complexContent >
    <xs:extension base="vstd:Standard" >
      <xs:sequence >
        <xs:element name="interface" type="vr:Interface" minOccurs="0"
                    maxOccurs="unbounded" />
      </sequence>
    </extension>
  </complexContent>
</complexType>
```

**vstd:ServiceStandard Extension Metadata Elements**

| Element | Definition | |
|---------|------------|--|
| interface | *Value type:* | composite: `vr:Interface` |
| | *Semantic Meaning:* | an abstract description of one of the interfaces defined by this service standard. |
| | *Occurrences:* | optional; multiple occurrences allowed. |
| | *Comments:* | This element can provide details about the interface that apply to all implementations. Each interface element should specify a `role` attribute with a value starting with "std:", or, if there is only one standard child element, `role` is set to "std". |

Even though the `vr:Interface` type requires an `<accessURL>` child element, the `<interface>` element in a `vstd:ServiceStandard` is intended to describe a service in the abstract--i.e. without reference to a particular installation of the service. Consequently, the accessURL should contain a bogus URL; applications should not expect it to be resolvable.

An application can, in principle, get a complete machine-readable description of a particular instance of a standard service (to, say, create a GUI for that service on-the-fly) by combining the general description in the `vstd:ServiceStandard` record with the service resource description for the specific instance. The intended process for building that description is as follows:

1. The application obtains a VOResource resource record for the service instance (e.g. from a registry).
2. The application extracts the `standardID` attribute for the desired service capability.
3. The `standardID` is resolved (via a registry) to a `vstd:ServiceStandard` record for the service. This description would capture the required and optional (but standard) components of the service interface.
4. The specific instance's interface description is merged into the standard one. The service's support of optional components as well as its allowed customizations would override the generic description from the `vstd:ServiceStandard` record.

The so-called "simple" data access layer (DAL) services, such as the Simple Image Access services [SIA], are registered using the `vs:ParamHTTP` interface type [VDS] to describe its interface. This interface type allows one to list input parameters accepted by the service. Each parameter can be marked as *required*, *optional*, or *ignored*. Typically with DAL services, parameters defined as optional by the standard may be legally ignored by an implementation. Consequently, this specification recommends special instruction for listing and interpreting input parameteters in a `vstd:ServiceStandard` record when the interface is of type `vs:ParamHTTP`: parameters that can be optionally provided in a client's query but are ignorable by the implementation should be marked as *ignored*. Applications that consume such `vstd:ServiceStandard` records should thus interpret the parameters marked *ignored* as *optional* for use by clients and *ignorable* by implementations. This minimizes the list of parameters that the service provider must list in the registration of a particular service instance to the ones that are actually supported (i.e. not ignored): when the list service instance description is merged into the list from the `vstd:ServiceStandard` record (step 4 above), the result is an accurate list of the optional but supported and the ignored parameters for that service instance.

An example of an instance of a `vstd:ServiceStandard` record is shown in Appendix B. It describes the Simple Image Access Specification [SIA] and in particular illustrates the recommended way to list input parameters defined by the standard.

### 3.1.3. StandardKeyEnumeration

The `vstd:StandardKeyEnumeration` resource type is available for collecting definitions of related, standard keys. Each key defined within this resource can then be referred to by a unique IVOA Identifier URI (see section 3.2). To support this, the `vstd:StandardKeyEnumeration` resource simply adds the `<key>` element to the standard core metadata.

| vstd:StandardKeyEnumeration Type Schema Definition |
|---|
| ```
<xs:complexType name="StandardKeyEnumeration" >
  <xs:complexContent >
    <xs:extension base="vr:Resource" >
      <xs:sequence >
        <xs:element name="key" type="vstd:StandardKey" maxOccurs="unbounded"
                    minOccurs="1" />
      </sequence>
    </extension>
  </complexContent>
</complexType>
``` |

| vstd:StandardKeyEnumeration Extension Metadata Elements | |
|---|---|
| **Element** | **Definition** |
| key | *Value type:*     composite: `vstd:StandardKey` |
| | *Semantic Meaning:* the name and definition of a key--a named concept, feature, or property. |
| | *Occurrences:*     required; multiple occurrences allowed. |

The contents of the `<key>` element is described in the next section.

## 3.2. Defining Keys: StandardKey and StandardKeyURI

The `vstd:StandardKey` type provides the means to define keys (as defined in section 2.3) within a VOResource record.

| vstd:StandardKey Type Schema Definition |
|---|
| ```
<xs:complexType name="StandardKey" >
  <xs:sequence >
    <xs:element name="name" type="vstd:fragment" />
    <xs:element name="description" type="xs:token" />
  </sequence>
</complexType>

<xs:simpleType name="fragment" >
  <xs:restriction base="xs:string" >
    <xs:pattern value="([A-Za-z0-9;/\?:@&=\+$,\-_.!~\*'\(\)]|%[A-Fa-f0-9]{2})+" />
  </restriction>
</simpleType>
``` |

| vstd:StandardKey Metadata Elements | | |
|---|---|---|
| **Element** | **Definition** | |
| name | *Value type:* | string with a form restricted to a legal URI fragment [URI]. |
| | *Semantic Meaning:* | The property identifier which would appear as the fragment (string after the pound sign, #) in an IVOA identifier. |
| | *Occurrences:* | required |
| | *Comments:* | Note that fragments may not include a pound (#) sign. |
| description | *Value type:* | string: `xs:token` |
| | *Semantic Meaning:* | A human-readable definition of this property. |
| | *Occurrences:* | required |
| | *Comments:* | Note that while a lengthy definition and other comments can be included in the body of this element, the `xs:token` type does not support multiple paragraphs. |

Defining a key via a `<key>` element within a VOResource record implies the definition of a unique URI formed according to the syntax described in section 2.3 that represents the semantics given by the value of the `<description>` child element. Because the URI must be globally unique, the key name (given by the `<name>` child element) must be unique within the VOResource record.

Though it is not needed by StandardsRegExt resource records, the StandardsRegExt schema further defines a convenience type, `vstd:StandardKeyURI`, which defines the legal pattern for a full standard key identifier (as defined in section 2.3). Applications that wish to use XML Schema to validate the form of a key URI may import the StandardsRegExt schema and use this type.

> **Note:**
> It is worth noting that just using or otherwise referencing a standard key URI in an application does not require importing the StandardsRegExt nor need there be any reference to the StandardsRegExt namespace. The role of the StandardsRegExt schema is simply to provide a way of documenting the definitions in a VOResource record. Thus, an application may dereference the URI for display or user help purposes; however, dereferencing is not necessary to use the URI.

# Appendix A: The complete StandardsRegExt Schema

The schema is included here for completeness - the definitive source of the schema is at
http://www.ivoa.net/xml/StandardsRegExt/v1.0.

---

**The Complete StandardsRegExt Schema**

```xml
<?xml version="1.0" encoding="UTF-8"?>
<xs:schema targetNamespace="http://www.ivoa.net/xml/StandardsRegExt/v1.0"
           xmlns:xs="http://www.w3.org/2001/XMLSchema"
           xmlns:vr="http://www.ivoa.net/xml/VOResource/v1.0"
           xmlns:vstd="http://www.ivoa.net/xml/StandardsRegExt/v1.0"
           xmlns:vm="http://www.ivoa.net/xml/VOMetadata/v0.1"
           elementFormDefault="unqualified" attributeFormDefault="unqualified"
           version="1.0" >

    <xs:annotation>
        <xs:appinfo>
            <vm:schemaName>StandardsRegExt</vm:schemaName>
            <vm:schemaPrefix>xs</vm:schemaPrefix>
            <vm:targetPrefix>vstd</vm:targetPrefix>
        </xs:appinfo>
        <xs:documentation>
            This is a core schema for describing IVOA Standards themselves
        </xs:documentation>
    </xs:annotation>

    <xs:import namespace="http://www.ivoa.net/xml/VOResource/v1.0"
               schemaLocation="http://www.ivoa.net/xml/VOResource/v1.0"/>

    <xs:complexType name="Standard">
        <xs:annotation>
            <xs:documentation>
              a description of a standard specification.
            </xs:documentation>
            <xs:documentation>
              This typically refers to an IVOA standard but is not
              limited to such.
            </xs:documentation>
        </xs:annotation>

        <xs:complexContent>
            <xs:extension base="vr:Resource">
                <xs:sequence>

                    <xs:element name="endorsedVersion" type="vstd:EndorsedVersion"
                                maxOccurs="unbounded">
                        <xs:annotation>
                            <xs:documentation>
                              the version of the standard that is recommended for use.
                            </xs:documentation>
                            <xs:documentation>
                              More than one version can be listed, indicating
                              that any of these versions are recognized as
                              acceptable for use.
                            </xs:documentation>
                        </xs:annotation>
                    </xs:element>

                    <xs:element name="schema" type="vstd:Schema"
                                minOccurs="0" maxOccurs="unbounded">
                        <xs:annotation>
                            <xs:documentation>
                              a description and pointer to a schema document
                              defined by this standard.
                            </xs:documentation>
                            <xs:documentation>
                              This is most typically an XML Schema, but it need
                              not be strictly.
                            </xs:documentation>
                        </xs:annotation>
                    </xs:element>

                    <xs:element name="deprecated" type="xs:token" minOccurs="0">
                        <xs:annotation>
                            <xs:documentation>
                                when present, this element indicates that all
```

```
                          versions of the standard are considered
                          deprecated by the publisher.  The value is a
                          human-readable explanation for the designation.
                    </xs:documentation>
                    <xs:documentation>
                       The explanation should indicate if another
                       standard should be preferred.
                    </xs:documentation>
                 </xs:annotation>
              </xs:element>

              <xs:element name="key" type="vstd:StandardKey"
                           minOccurs="0" maxOccurs="unbounded">
                 <xs:annotation>
                    <xs:documentation>
                       a defined key associated with this standard.
                    </xs:documentation>
                 </xs:annotation>
              </xs:element>

           </xs:sequence>
        </xs:extension>
     </xs:complexContent>
</xs:complexType>

<xs:complexType name="EndorsedVersion">
   <xs:simpleContent>
      <xs:extension base="xs:string">
         <xs:attribute name="status" default="n/a">
            <xs:annotation>
               <xs:documentation>
                  the IVOA status level of this version of the standard.
               </xs:documentation>
               <xs:documentation>
                  For values of "rec", "pr", "wd", and "note", the
                  record's referenceURL element should point to the
                  official specification document in the IVOA Document
                  Repository; if the document does not appear there,
                  these values should not be used.
               </xs:documentation>
            </xs:annotation>

            <xs:simpleType>
               <xs:restriction base="xs:string">
                  <xs:enumeration value="rec">
                     <xs:annotation>
                        <xs:documentation>
                           an IVOA Recommendation
                        </xs:documentation>
                     </xs:annotation>
                  </xs:enumeration>
                  <xs:enumeration value="pr">
                     <xs:annotation>
                        <xs:documentation>
                           an IVOA Proposed Recommendation
                        </xs:documentation>
                     </xs:annotation>
                  </xs:enumeration>
                  <xs:enumeration value="wd">
                     <xs:annotation>
                        <xs:documentation>
                           an IVOA Working Draft
                        </xs:documentation>
                     </xs:annotation>
                  </xs:enumeration>
                  <xs:enumeration value="iwd">
                     <xs:annotation>
                        <xs:documentation>
                           an internal Working Draft of an IVOA Working Group
                        </xs:documentation>
                     </xs:annotation>
                  </xs:enumeration>
                  <xs:enumeration value="note">
                     <xs:annotation>
                        <xs:documentation>
                           a published IVOA Note
                        </xs:documentation>
                     </xs:annotation>
                  </xs:enumeration>
                  <xs:enumeration value="n/a">
```

```
                    <xs:annotation>
                        <xs:documentation>
                            not an IVOA standard or protostandard at
                            this time.
                        </xs:documentation>
                    </xs:annotation>
                </xs:enumeration>
            </xs:restriction>
          </xs:simpleType>
        </xs:attribute>

        <xs:attribute name="use">
          <xs:annotation>
            <xs:documentation>
              A designation of preference for the version compared
              to other versions in use.
            </xs:documentation>
          </xs:annotation>

          <xs:simpleType>
            <xs:restriction base="xs:string">
              <xs:enumeration value="preferred">
                <xs:annotation>
                    <xs:documentation>
                        the most preferred version.
                    </xs:documentation>
                    <xs:documentation>
                        Only one version should have this designation.
                    </xs:documentation>
                </xs:annotation>
              </xs:enumeration>
              <xs:enumeration value="deprecated">
                <xs:annotation>
                    <xs:documentation>
                        a version whose use is now discouraged
                        because a newer version is preferred.
                    </xs:documentation>
                </xs:annotation>
              </xs:enumeration>
            </xs:restriction>
          </xs:simpleType>
        </xs:attribute>

      </xs:extension>
    </xs:simpleContent>
</xs:complexType>

<xs:complexType name="Schema">
    <xs:annotation>
        <xs:documentation>
          a description of a schema definition
        </xs:documentation>
    </xs:annotation>

    <xs:sequence>
      <xs:element name="location" type="xs:anyURI"
                  minOccurs="1" maxOccurs="1">
        <xs:annotation>
            <xs:documentation>
              A URL pointing to a document that formally defines
              the schema.
            </xs:documentation>
            <xs:documentation>
              The document should be in a machine-parsable format
              when applicable.  For example, when refering to an
              XML schema, the document should be an XML Schema or
              similar document that can be used to validate an
              instance document.
            </xs:documentation>
        </xs:annotation>
      </xs:element>
      <xs:element name="description" type="xs:token"
                  minOccurs="0" maxOccurs="1" >
        <xs:annotation>
            <xs:documentation>
              A human-readable description of what the schema
              defines or is used for.
            </xs:documentation>
            <xs:documentation>
              A brief description--e.g. one statement--is
```

```
                        recommended for display purposes.
                    </xs:documentation>
                </xs:annotation>
            </xs:element>
            <xs:element name="example" type="xs:anyURI"
                        minOccurs="0" maxOccurs="unbounded">
                <xs:annotation>
                    <xs:documentation>
                        A URL pointing to a sample document that illustrates
                        the use of the schema.
                    </xs:documentation>
                    <xs:documentation>
                        When applicable (e.g. XML), the document should be
                        in the format defined by the schema document.
                    </xs:documentation>
                </xs:annotation>
            </xs:element>
        </xs:sequence>

        <xs:attribute name="namespace" type="xs:token" use="required">
            <xs:annotation>
                <xs:documentation>
                    an identifier for the schema being described.  Each instance
                    of this attribute must be unique within the resourse description.
                </xs:documentation>
                <xs:documentation>
                    For XML schemas, this should be the schema's namespace URI.
                    Otherwise, it should be a unique label to distinguish it from
                    other schemas described in the same resource description.
                </xs:documentation>
            </xs:annotation>
        </xs:attribute>
    </xs:complexType>

    <xs:complexType name="ServiceStandard">
        <xs:annotation>
            <xs:documentation>
                a description of a standard service protocol.
            </xs:documentation>
            <xs:documentation>
                This typically refers to an IVOA standard but is not
                limited to such.
            </xs:documentation>
        </xs:annotation>

        <xs:complexContent>
            <xs:extension base="vstd:Standard">
                <xs:sequence>
                    <xs:element name="interface" type="vr:Interface"
                                minOccurs="0" maxOccurs="unbounded">
                        <xs:annotation>
                            <xs:documentation>
                                an abstract description of one of the interfaces defined
                                by this service standard.
                            </xs:documentation>
                            <xs:documentation>
                                This element can provide details about the interface
                                that apply to all implementations.  Each interface
                                element should specify a role with a value starting
                                with "std:" or, if there is only one standard interface,
                                is equal to "std".
                            </xs:documentation>
                            <xs:documentation>
                                Applications that, for example, wish to build a GUI
                                to the service on-the-fly would first access this generic
                                description.  Site-specific variations, such
                                as supported optional arguments or site specific
                                arguments, would be given in the actual implemented
                                service description and overrides any common information
                                found in this generic description.  This generic interface
                                description can be matched with the site-specific one
                                using the role attribute.
                            </xs:documentation>
                            <xs:documentation>
                                Even though the Interface type requires an
                                accessURL child element, this element is
                                intended to describe a service in the
                                abstract--i.e. without reference to a particular
                                installation of the service.  Consequently,
                                the accessURL may contain a bogus URL;
```

```
                              applications should not expect it to be resolvable.
                        </xs:documentation>
                    </xs:annotation>
                </xs:element>
            </xs:sequence>
        </xs:extension>
    </xs:complexContent>
</xs:complexType>

<xs:complexType name="StandardKeyEnumeration">
    <xs:annotation>
        <xs:documentation>
            A registered set of related keys.  Each key can be
            uniquely identified by combining the IVOA identifier of
            this resource with the key name separated by the URI
            fragment delimiter, #, as in: ivoa-identifier#key-name
        </xs:documentation>
    </xs:annotation>
    <xs:complexContent>
        <xs:extension base="vr:Resource">
            <xs:sequence>
                <xs:element name="key" type="vstd:StandardKey"
                            maxOccurs="unbounded" minOccurs="1">
                    <xs:annotation>
                        <xs:documentation>
                            the name and definition of a key--a named concept,
                            feature, or property.
                        </xs:documentation>
                    </xs:annotation>
                </xs:element>
            </xs:sequence>
        </xs:extension>
    </xs:complexContent>
</xs:complexType>

<xs:complexType name="StandardKey">
    <xs:annotation>
        <xs:documentation>
            The name and definition of a key--a named concept,
            feature, or property.
        </xs:documentation>
        <xs:documentation>
            This key can be identified via an IVOA identifier
            of the form ivo://authority/resource#name where name is
            the value of the child name element.
        </xs:documentation>
        <xs:documentation>
            This type can be extended if the key has
            other metadata associated with it.
        </xs:documentation>
    </xs:annotation>

    <xs:sequence>
        <xs:element name="name" type="vstd:fragment">
            <xs:annotation>
                <xs:documentation>
                    The property identifier which would appear as the
                    fragment (string after the pound sign, #) in an IVOA
                    identifier.
                </xs:documentation>
            </xs:annotation>
        </xs:element>
        <xs:element name="description" type="xs:token">
            <xs:annotation>
                <xs:documentation>
                    A human-readable definition of this property.
                </xs:documentation>
            </xs:annotation>
        </xs:element>
    </xs:sequence>
</xs:complexType>

<xs:simpleType name="StandardKeyURI">
    <xs:annotation>
        <xs:documentation>
            reference that forces an IVOA ID with a # fragment part
            appended to match the standard pattern for registering
            enumeration values with a vstd:StandardKeyList
        </xs:documentation>
    </xs:annotation>
```

```
        <xs:restriction base="xs:anyURI">
            <xs:pattern value="ivo://[\w\d][\w\d\-_\.!~\*'\(\)\+=]{2,}(/[\w\d\-_\.!~\*'\(\)\+=]+(/[\w\d
        </xs:restriction>
    </xs:simpleType>

    <xs:simpleType name="fragment">
        <xs:annotation>
            <xs:documentation>
                the allowed characters for a fragment identifier taken
                from rfc2396
            </xs:documentation>
        </xs:annotation>
        <xs:restriction base="xs:string">
            <xs:pattern
                value="([A-Za-z0-9;/\?:@&=\+$,\-_\.!~\*'\(\)]|%[A-Fa-f0-9]{2})+"/>
        </xs:restriction>
    </xs:simpleType>
</xs:schema>
```

# Appendix B: A Sample Record

This example shows how one can describe an IVOA standard, the Simple Image Access Protocol. It includes a description of the input parameters defined in the specification.

**A sample record describing the SIA standard**

```
><?xml version="1.0" encoding="UTF-8"?>
<resource xsi:type="vstd:ServiceStandard" status="active"
          created="2000-01-01T09:00:00" updated="2000-01-01T09:00:00"
          xmlns:vstd="http://www.ivoa.net/xml/StandardsRegExt/v1.0"
          xmlns:vs="http://www.ivoa.net/xml/VODataService/v1.1"
          xmlns:xsi="http://www.w3.org/2001/XMLSchema-instance"
          xsi:schemaLocation="http://www.ivoa.net/xml/VOResource/v1.0
                              VOResource-v1.0.xsd
                              http://www.ivoa.net/xml/VODataService/v1.1
                              VODataService-v1.1.xsd
                              http://www.ivoa.net/xml/StandardsRegExt/v1.0
                              StandardsRegExt-v1.0.xsd">

    <title>  Simple Image Access Protocol   </title>
    <shortName>  SIA   </shortName>
    <identifier>   ivo://ivoa.net/std/SIA    </identifier>

    <curation>
        <publisher ivo-id="ivo://ivoa.net/IVOA">
          International Virtual Observatory Alliance
        </publisher>
        <creator>
          <name>  Doug Tody   </name>
        </creator>
        <creator>
          <name>  Ray Plante   </name>
        </creator>
        <date>  2004-05-24   </date>
        <contact>
          <name>  Data Access Layer Working Group   </name>
          <email>  dal@ivoa.net   </email>
        </contact>
    </curation>

    <content>
        <subject>  software standard   </subject>
        <subject>  virtual observatory   </subject>
        <description>
          The Simple Image Access Protocol is a protocol for retrieving
          image data from a variety of astronomical image repositories
          through a uniform interface. The interface is meant to be
          reasonably simple to implement by service providers. A query
          defining a rectangular region on the sky is used to query for
          candidate images. The service returns a list of candidate
          images formatted as a VOTable. For each candidate image an
          access reference URL may be used to retrieve the image. Images
          may be returned in a variety of formats including FITS and
          various graphics formats. Referenced images are often computed
          on the fly, e.g., as cutouts from larger images.
        </description>
```

```xml
  <referenceURL>
     http://www.ivoa.net/Documents/latest/SIA.html
  </referenceURL>
  <type>   Other   </type>
  <contentLevel>   Research   </contentLevel>
</content>

<endorsedVersion status="rec"> 1.0 </endorsedVersion>

<interface xsi:type="vs:ParamHTTP" role="std" version="1.0">
   <accessURL use="base"> http://sample.org/cgi-bin/sia </accessURL>
   <queryType>GET</queryType>
   <resultType>text/xml+votable</resultType>

   <!--
     -  These are the standard input parameters defined in the
     -  SIA spec
     -->
   <param use="required">
     <name>POS</name>
     <description>
        Search Position in the form "ra,dec" where ra and dec are given
        in decimal degrees in the ICRS coordinate system.
     </description>
     <unit>degrees</unit>
     <dataType arraysize="2">real</dataType>
   </param>
   <param use="required">
     <name>SIZE</name>
     <description>
        Size of search region in the RA and Dec. directions in decimal
        degrees.
     </description>
     <unit>degrees</unit>
     <dataType arraysize="2">real</dataType>
   </param>
   <param use="optional">
     <name>FORMAT</name>
     <description>Requested format of images.</description>
     <dataType>string</dataType>
   </param>
   <param use="optional">
     <name>INTERSECT</name>
     <description>
        Choice of intersection of matched images with the region of
        interest.
     </description>
     <dataType>string</dataType>
   </param>
   <param use="ignored">
     <name>NAXIS</name>
     <description>
        The number of pixels desired along each axis
     </description>
     <dataType arraysize="2">integer</dataType>
   </param>
   <param use="ignored">
     <name>CFRAME</name>
     <description>
        the coordinate frame to impose on the image.
     </description>
     <dataType>string</dataType>
   </param>
   <param use="ignored">
     <name>EQUINOX</name>
     <description>
        the epoch of the mean equator and equinox for the specified
        coordinate system reference frame (CFRAME). coordinate frame to
        impose on the image.
     </description>
     <dataType>string</dataType>
   </param>
   <param use="ignored">
     <name>CRPIX</name>
     <description>
        the pixel position to locate the reference position in the output
        image.
     </description>
     <dataType arraysize="2">real</dataType>
   </param>
```

```
        <param use="ignored">
          <name>CRVAL</name>
          <description>
            the world coordinates of the reference position in the output
            image in decimal degrees
          </description>
          <dataType arraysize="2">real</dataType>
        </param>
        <param use="ignored">
          <name>CDELT</name>
          <description>
            the scale of the output image in decimal degrees per pixel
          </description>
          <dataType arraysize="2">real</dataType>
        </param>
        <param use="ignored">
          <name>ROTANG</name>
          <description>
            the rotation angle to put into the output image's coordinate system
          </description>
          <dataType arraysize="2">real</dataType>
        </param>
        <param use="ignored">
          <name>PROJ</name>
          <description>
            the projection to impose in the construction of the output image
          </description>
          <dataType>string</dataType>
        </param>
        <param use="ignored">
          <name>VERB</name>
          <description>
            the level of verbosity in the output.
          </description>
          <dataType>string</dataType>
        </param>
      </interface>
    </resource>
```

## Appendix C: Change History

**Changes since PR-v1.0 20120217:**

- none other than date and status.

**Changes since PR-v1.0 20120213:**

- added the `<schema>` element to `vstd:Standard`
- updated example in section 3.1.1

**Changes since PR-v1.0 20111017:**

- added Note box to section 2.2 that recommends against using `vstd:StandardKeyEnumeration` to describe keys when they are defined by an IVOA standard.
- added statement in section 2.3 highlighting that # signs are not allowed in key names.
- added `iwd` and `note` as allowed values for `vstd:EndorsedVersion`'s `status` attribute.
- converted a Note box in section 3.1.2 to a normative paragraph that recommends listing optional `ParamHTTP` parameters as ignored. Note that there is a related error in the VODataService standard [VDS]: while the `ignored` value is defined in the schema, it is not included in the document text.
- added sample `vstd:ServiceStandard` instance record back in as Appendix B
- added references to TAP and SIA
- fixed various grammatical typos.

**Changes since PR-v1.0 20110921:**

- In `<endorsedVersion>`, changed "prop" to "pr".
- various typo corrections

**Changes since PR-v1.0 20110316:**

- Corrected ampersand representation in schema listing
- Various typo corrections and clarifications

**Changes since WD-v1.0 20100519:**

- Prepped for PR
- improved discussion of example in section 2
- Added standard architecture sub-section
- updated in-lined schema (App. 1)

**Changes since WD-v1.0 20100514:**

- short name changed from VOStandard to StandardsRegExt

**Changes since WD-v0.4:**

- removed App. B. (Sample instance) as examples appear throughout the document.

**[TAP]** Dowler, Patric, Rixon, Guy, and Tody, Doug 2010, *Table Access Protocol, Version 1.0*, IVOA Recommendation, `http://www.ivoa.net/Documents/TAP/20100327/`.

**[VDS]** Plante, R., Stébé, A. Benson, K., Dowler, P. Graham, M., Greene, G., Harrison, P., Lemson, G., Linde, T., Rixon, G., Stébé, A. 2010, *VODataService: a VOResource Schema Extension for Describing Collections and Services*, v1.1, IVOA Recommendation, `http://www.ivoa.net/Documents/VODataService/20101202/`